\documentclass[sigconf]{acmart}

\usepackage{booktabs} 

\setcopyright{rightsretained}

\acmDOI{10.475/123_4}

\acmISBN{123-4567-24-567/08/06}

\acmConference[SIGIR eCom 2018]{SIGIR 2018 Workshop on eCommerce}{July 2018}{ Ann Arbor, Michigan, USA}
\acmYear{2018}
\copyrightyear{2018}

\acmArticle{4}
\acmPrice{15.00}

\begin{document}
\title{Towards a simplified ontology for better e-commerce search}

\author{Aliasgar Kutiyanawala}
\affiliation{%
  \institution{Jet.com and Walmart Labs}
  \streetaddress{221 River Street}
  \city{Hoboken}
  \state{New Jersey}
  \postcode{07030}
}
\email{aliasgar@jet.com}

\author{Prateek Verma}
\affiliation{%
  \institution{Jet.com and Walmart Labs}
  \streetaddress{221 River Street}
  \city{Hoboken}
  \state{New Jersey}
  \postcode{07030}
}
\email{prateek.verma@jet.com}

\author{Zheng (John) Yan}
\affiliation{%
  \institution{Jet.com and Walmart Labs}
  \streetaddress{221 River Street}
  \city{Hoboken}
  \state{New Jersey}
  \postcode{07030}
}
\email{john@jet.com}

\begin{abstract}
Query Understanding is a semantic search method that can classify tokens in a customer's search query to entities like \emph{Product}, \emph{Brand}, etc. This method can overcome the limitations of bag-of-words methods but requires an ontology. We show that current ontologies are not optimized for search and propose a simplified ontology framework designed specially for e-commerce search and retrieval. We also present three methods for automatically extracting product classes for the proposed ontology and compare their performance relative to each other.
\end{abstract}

%
%



\maketitle

\section{Introduction}

\noindent Search plays a vital part in any e-commerce site and a poor search system leads to customer frustration, which negatively affects both retention and conversion. Most e-commerce sites employ a bag-of-words search method which simply matches tokens in a customer's search query with relevant fields of SKUs (stock keeping unit but used here to describe any item sold by the site). This system is easy to implement specially with solutions like ElasticSearch~\cite{gormley2015elasticsearch} or Solr~\cite{grainger2014solr} but suffers from some significant drawbacks. This system is prone to returning irrelevant results because of its inability to understand what the customer is looking for. Consider an example search query: ``\texttt{men's black leather wallet}'' and let us assume that there are no SKUs that match this query exactly. The bag-of-words system will resort to a partial match and may return men's brown leather wallets (relevant) along with men's black leather belts (irrelevant). This problem is also evident when queries are similar in terms of words but actually relate to very different products. For example: the queries ``\texttt{camera with lens}'' and ``\texttt{lens for camera}'' may produce the same result if prepositions are ignored as stopwords. There are ways to augment the bag-of-words search system with a category pinpointing (or prediction) model, bigrams, etc. to improve the recall but this approach is still not very accurate. \\

\noindent A better approach to search is to use a query understanding system to understand the customer's search \emph{intent}~\cite{rose2004understanding, hu2009understanding}. One such method is to use a semantic annotation process described in~\cite{popov2004kim, glater2017intent} by using a well defined ontology to classify terms from the customer's search query. Going back to our previous example, if we were to classify tokens in ``\texttt{men's black leather wallets}'' as \quad \textbf{\texttt{men := Gender, black := Color, leather := Material, wallet := Product}}, it would allow the system to find exactly what the customer is looking for or make relevant substitutions if no such SKU could be found. This task is called \emph{Named Entity Recognition and Classification} (NERC), where entities like Product, Color, Material, etc. are recognized. Nadeau and Sekine~\cite{nadeau2007survey} provide an excellent overview of this field. We use Bi-directional LSTM-CRF as described by Lample et al. in~\cite{glample} for performing named entity recognition although other systems like GATE~\cite{Cunningham2002, cunningham2013getting} could also be used. The named entity tagger can accurately recognize the customer's intent by recognizing and classifying entities in the query as long as those entities are well defined. The problem is that most existing product ontologies are designed from a supply-side perspective and not from a search perspective.\\

\noindent We propose a simplified ontology framework specially designed from a search and retrieval perspective that contains three top-level \emph{concepts} - \emph{Product}, \emph{Brand} and \emph{Attribute} and five \emph{slots} (or \emph{properties}) - \emph{synonyms}, \emph{attributes}, \emph{primary\_attributes}, \emph{brands} and \emph{default\_product}. We show that these three entity classes along with five slots can provide relevant recall for a customer's search query. We further discuss this ontology in Section~\ref{sec:ontology} and provide insights into why each entity type and slot is necessary and how they help in retrieving relevant results.\\

\noindent Our contributions in this paper are creating a product ontology designed specifically for search and providing three methods to automatically extract \emph{Product concepts} for this ontology. We discuss this ontology in detail in Section~\ref{sec:ontology}. We provide an overview of the field of Ontology learning in Section~\ref{sec:ontology-learning} and discuss our methods to extract \emph{Product concepts} in Section~\ref{sec:methods}. Finally, we present our conclusions in Section~\ref{sec:conclusion}.

\section{Ontology}
\label{sec:ontology}
An ontology is a formal explicit description of a domain by identifying \emph{classes} (or \emph{concepts}), \emph{slots} and \emph{slot restrictions} between classes for a particular domain~\cite{noy2001ontology}. \emph{Classes} represent the main concepts in a domain and are related to physical objects in that domain, for example: \emph{TV}, \emph{Shirt} or \emph{Screen Size}. \emph{Slots} represents properties of objects and relationships between classes, for example: the slot \emph{attribute} links the classes \emph{TV} and \emph{Screen Size}. \emph{Slot Restrictions} impose restrictions on the values that can be taken by a slot, for example: we can impose the restriction that \emph{Screen Size} is a positive number.\\


\noindent Our goal is to design a product ontology that can be used for search purposes. This ontology must serve a dual purpose - we must be able to classify SKUs onto this ontology and secondly, the classes (and subclasses) in this ontology should serve as named entities for query-side named entity recognition and classification. There are many supply-side ontologies for e-commerce like ecl@ss, Harmonised System, NAICS/NAPCS, RosettaNet, etc.~\cite{ding2004role} but they tend to focus more on relationships between buyers and sellers. They tend to include slots (or properties) such as \emph{GLN of manufacturer}, \emph{GLN of supplier}, \emph{product article number of supplier}, etc. which are completely unnecessary for search purposes. These ontologies also have product types that are very complex, for example the entire phrase: \emph{``Shirts, underwear, men's and boys', cut and sewn from purchased fabric (except apparel contractors)''} is a product from NAICS. Such product types contain attributes (like men's, boy's, etc.) along with the most basic form of the product (shirt) and hence are not considered \emph{atomic}. The NERC system will have a lot of difficulty in using such \emph{non-atomic} products.

\begin{figure}[!htbp]
  \includegraphics[width=\linewidth]{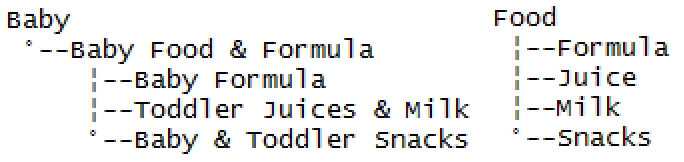}
  \caption{Comparison of catalog-side and search-side ontologies.}
  \label{fig:compare-ontologies}
\end{figure}

\noindent Work has also been done on ontologies that are focused more on the catalog side~\cite{charron2016extracting, lee2006building}. Catalog-side ontologies are closer to search-side ontologies as compared to supply-side ontologies but are still not perfectly aligned with a search perspective. Consider Figure~\ref{fig:compare-ontologies}, which shows a snippet of \emph{Product} classes from two ontologies - a catalog-side ontology on the left and a search-side ontology on the right. There are three main differences between them. The first difference is that the ontology on the left does not have a \emph{``is a''} relationship between classes and subclasses. For example: \emph{Baby food and formula} is not a \emph{Baby}. The ontology on the left tries to classify items by their intended use case but ontology on the right classifies items according to what they represent. The second difference is that the ontology on the left contains combo products like \emph{Toddler Juices and Milk}, which makes it difficult to know if a SKU classified to this product type is a \emph{Juice} or \emph{Milk}. The third difference is that the ontology on the left contains non-atomic entries like \emph{Baby and Toddler Snacks}, which should just be simplified to \emph{Snacks} as it makes it very easy for the NERC system to identify products in queries like ``\texttt{snacks for baby}.''\\

\noindent Our ontology contains a restriction that requires all classes (and subclasses) to be as \emph{atomic} as possible to improve recall. We define an \emph{atomic} entity as an irreducible unit that describes a concept. It also places a \emph{``is-a''} requirement on all subclasses for a given class. Finally, it tries to avoid combo classes unless they are sold as a set (\emph{dining sets} that must contain both \emph{table} and \emph{chairs}). This requirement keeps the ontology simple and flexible. The following sections describe the classes and slots in our ontology in greater detail. 

\subsection{Product}
\noindent A \emph{Product} is defined as the \emph{atomic} phrase that describes what the customer is looking for. Consider an example, ``\texttt{white chair with ottoman}''. Here, the customer is looking to buy a \emph{chair}. It is preferable if the chair is white in color and comes with an ottoman but these requirements are secondary to the primary requirement of it being a \emph{chair}. If such a chair is not available, the customer is more likely to buy a chair in a different color or one that does not come with an ottoman but is less likely to buy a white sofa with ottoman even though it satisfies two requirements out of three. Any specialized product type like \emph{folding chair} must be stripped down to its most basic form \emph{chair}. There are exceptions to this rule, for example,  a \emph{bar stool} is a specialized type of stool and ideally we should strip it down to its most basic form \emph{stool} but many customers use the term ``\texttt{barstool}'' (single term without spaces) to describe it. The NERC system has to be able to classify this term to a product and hence we include the term ``\texttt{barstool}'' as a Product in our ontology with the synonym ``\texttt{bar stool}''. The class \emph{barstool} is a sub-class of the class \emph{stool} because every barstool is ultimately a stool. This parent-child relationship also helps during recall because if the customer searches for ``\texttt{stool}'', the search system will include all stools including barstools in the recall. It should be noted that \emph{atomic} does not imply a single-word token because many multi-word tokens like \emph{air conditioner} and \emph{onion rings} are \emph{atomic}. We use a combination of our query and SKU understanding systems along with user data to provide suggestions for parent-child relationships and synonyms (or variations). However, describing this method is beyond the scope of this paper.

\subsection{Attribute}
\label{sec:attribute}
\noindent\emph{Attribute} is defined as an \emph{atomic} phrase that provides more information about an item. Consider an example ``\texttt{white wooden folding adirondack chair}''. Here, we classify the term {\emph{chair}} as a \emph{Product} and we can classify the remaining terms (white, wooden, folding and adirondack) as \emph{Attributes}. This gives us a lot of flexibility during recall. Initially, the search system can restrict the recall by filtering out any SKUs that do not match the product type and then boost SKUs by the number of matching attributes. In case of our example, we would restrict the recall to be chairs of all types and then boost those SKUs that match the attributes (white, wooden, folding and adirondack). A SKU that matches all attributes will have a higher score (and placed on top of the recall) than those that match fewer attributes.\\

\noindent Attributes can be subclassed as \emph{Color}, \emph{Material}, \emph{SleeveType}, etc. depending on the category. We found that only a subset of Attributes are relevant for search purposes. An attribute like \emph{Country of Manufacture} may be a valid subclass but it can be argued that it is not very important for search purposes. Since our aim is to create a \emph{simplified} ontology for search, we restrict attribute subclasses to what is actually important for search. This makes the system much more maintainable. The range of most attributes are values from an enumerated set but some attributes like \emph{Screen Size} may have numeric values along with a unit of measurement like \emph{inches}, \emph{cm}, etc. Such numeric values can be normalized using simple rules (\emph{1 inch = 2.54 cm}) so that more relevant SKUs can be recalled for a given query even if they have units from different measurement systems. It is not necessary that numeric values in the query and SKU to match exactly. We compute the difference between corresponding numeric values of the query and SKU and apply a boost that is inversely proportion to the difference. For example, a query: ``\texttt{45 inch tv}'' will match SKUs for \emph{43 inch TVs} (higher boost) as well as \emph{49 inch TVs} (lower boost)\\

\subsection{Brand}
\label{sec:brand}
A \emph{Brand} is defined as a phrase that provides more information about the manufacturer of the item. \emph{Samsung}, \emph{Calvin Klein}, etc. are examples of brands. Brands are important because they capture information about the preferences of the customer but are not essential in defining the recall. The search system tries to honor the customer's preference regarding the brand by boosting SKUs that match the brand specified in the query. This scheme ensures that the search result includes SKUs from other brands albeit at a lower position compared to SKUs that match the brand in the query.\\

\noindent We observed that in some cases customers tend to use the brand name as a synonym for a product, for example, ``\texttt{q-tips}'' to denote cotton swabs and ``\texttt{kleenex}'' to denote tissues. This type of behavior is common for a subset of brands that have high brand equity and are taken to represent the product itself. We wanted to respect the customer's preferences while still providing them with a wide range of similar products from other brands and so we introduced the \emph{default\_product} relation, which maps these finite subsets of brands with their default \emph{Product} nodes. This relation then allows the NERC system to map the query ``\texttt{kleenex}'' to \textbf{\texttt{kleenex := Brand, tissues := Product}} and have the flexibility to present relevant SKUs from other brands at a lower position in the recall.

\noindent Currently, we do not support a parent-child relationship between brands (for example: \emph{Nike}) and sub-brands (for example: \emph{Nike Air}). and treat each sub-brand as a variation of the original brand. 

\subsection{Slots}
We propose five slots or properties - \emph{synonyms}, \emph{attributes}, \emph{primary\_attributes}, \emph{brand} and \emph{default\_product} and show how they can be used to recall relevant SKUs for a given query. The \emph{synonyms} slot indicates synonyms of a given class and are typically used to address alternate phrases used to describe the same item. The \emph{synonyms} slot exists for all classes in our ontology.\\

\noindent The \emph{attributes} slot has the \emph{Product} class as its domain and the \emph{Attributes} class as the range. It helps in specifying all relevant attributes for a given SKU. Since we insist on \emph{atomic} products, this slot helps us in distinguishing relevant SKUs from irrelevant SKUs in the recall. Consider the two queries ``\texttt{Dining Chair}'' and ``\texttt{Outdoor Chair}'', which refer to two very \emph{different} products even though they are both \emph{chairs}. The NERC system is able to extract the attributes \emph{dining} and \emph{outdoor} for those two queries and is able to boost SKUs that match these attributes to the top of the recall. Thus, the customer is presented with relevant SKUs in each case even though the product type of both queries is the same.\\

\noindent Consider a search query ``\texttt{cotton shirt}'', where the NERC system is able to extract the material \emph{cotton}. As discussed previously, the system will retrieve all shirts and automatically boost cotton shirts so that they appear the top of the recall. Let us assume that there are two SKUs - one shirt made of 100\% cotton and the other shirt made out of 95\% polyester and only 5\% cotton. If there is no notion of \emph{primary\_attributes} both SKUs will receive the same attribute boost and will be considered equally relevant. The \emph{primary\_attributes} is a special slot that maps a \emph{Product} with a single \emph{Material} or \emph{Color} subclass. In case of the previous example the \emph{primary\_attribute} will point to \textbf{\texttt{cotton := Material}} for the first SKU and \textbf{\texttt{polyester := Material}} for the second. This slot helps increase relevancy by boosting only SKUs that match the corresponding primary color or material.\\

\noindent The \emph{Brands} slot has the \emph{Product} class as the domain and the \emph{Brands} class as the range. It defines the manufacturer for a given SKU. As mentioned previously, the \emph{default\_product} slot helps in assigning a product to a small set of brands like \emph{Kleenex} that are used synonymously with products. Both slots help increase relevancy by boosting all SKUs that match the extracted brand from the query but without sacrificing the ability to show SKUs from other brands at lower positions on the search page.\\

\noindent Our current implementation of ranking SKUs is rather simple - providing fixed boosts when products, brand and attributes from the query match products, brands and attributes in the SKU. In future, we will use these matches in conjunction with a ranking model to further improve relevancy.

\section{Ontology Learning}
\label{sec:ontology-learning}
\noindent The task of building an ontology is a time consuming and expensive task and usually involves a domain expert. Techniques that support ontology engineering and reduce the cost of building and maintaining ontology are required to ensure that this task is scalable. Ontology learning can be thought of as data driven methods that support building ontologies by deriving classes and meaningful relations between them. Petucci et al~\cite{petrucci2016ontology} formulate the problem of ontology learning from natural language as transductive reasoning task that learns to convert natural language to a logic based specification. It breaks down the problem into two tasks - sentence transduction phase and sentence tagging phase. It uses RNN for sentence tagging and RNN Encoder-Decoder model for sentence transduction.\\

\begin{figure}[!htbp]
  \includegraphics[width=\linewidth]{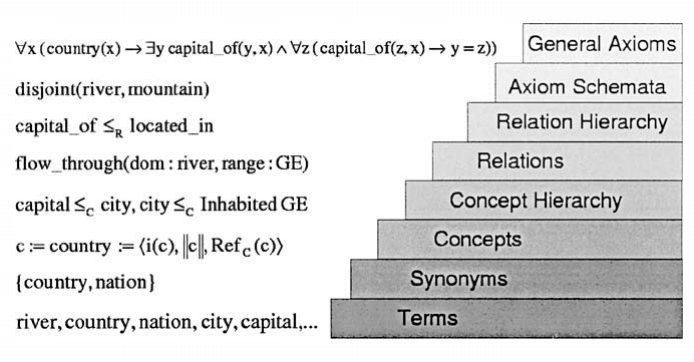}
  \caption{Ontology Learning Layer Cake.}
  \label{fig:layer-cake}
\end{figure}

\noindent Figure~\ref{fig:layer-cake} shows the concept of \emph{ontology learning layer cake}, which was introduced by Cimiano et. al~\cite{Cimiano:2006:OLP:1177318} and further discussed in~\cite{maedche2004ontology}. The layers focus on ontology learning and show dependencies between various tasks in the ontology learning system. The layers are designed such that results of lower layers serve as inputs to higher layers.\\

\noindent The \emph{term extraction layer} is the lowest layer in the cake. It aims to learn the relevant terminology of the domain. A naive approach is to just use term frequencies assuming that relevant concepts are also most frequent. However, other sophisticated methods like TF-IDF~\cite{zhang2008comparative} or C-value/NC-value measure proposed in~\cite{frantzi1999c} can also be used.\\

\noindent The next layer is the \emph{synonym extraction layer}, which deals with extracting synonyms for the terms identified in the previous layer. Synonyms can be extracted using a distributional representation of words, which claim that similar words share similar contexts~\cite{wohlgenannt2016using}. Semantic relatedness using wordnet or Wikipedia categories can be used as well~\cite{gabrilovich2007computing}.\\

\noindent The third layer is the \emph{concept formation layer}, which provides a definition of concepts, their extension and the lexical signs which are used to refer to them. The fourth layer is the \emph{concept hierarchy layer}, which deals with inducing, extending and refining the ontology hierarchy. This task can be accomplished by methods like matching lexico-syntactic patterns as demonstrated by Hearst in~\cite{hearst1992automatic}, clustering different objects based on their feature vectors and using phrase analysis i.e., making use of internal structure of noun phrases to discover taxonomic relations~\cite{sanchez2005web}.\\

\noindent The fifth and sixth layers deal with \emph{Relations}, which is the task of learning relation labels (or identifiers) as well as their corresponding domain and range. Some common methods include finding co-occurrence between words as proposed by Madche~\cite{maedche2000discovering}.\\

\noindent The last two layers are \emph{Axiom Schemata} and \emph{General Axioms}, which are related to rules and axioms. These two layers deal with transformation of natural language definitions into OWL Description Logic axioms, and building a domain specific ontology by pruning an existing general ontology using the given corpus~\cite{buitelaar2001ranking}.\\

\noindent Since we do not deal with axioms, the last two layers of the ontology learning cake are not relevant to our task. The first three layers require most manual effort and are most time consuming for our task. This paper describes three methods that can automatically derive terms and \emph{Product} concepts, which correspond to the first and the third layer in the ontology learning layer cake. Ontology creation cannot be fully automated and our methods produce candidates for manual review, greatly decreasing the time required for ontology development. These methods do not address the problem of synonym resolution but other methods that use click logs on top of an existing ontology can help with the second layer. Unfortunately, describing this method is beyond the scope of this paper.

\section{Automatically extracting Product Entities}
\label{sec:methods}
\noindent We describe three methods (Token Graph Method, Augmented Graph Method and LSTM-CRF method) that can be used to automatically extract atomic Product entities from a customer's search query. Two of these methods may also be extended to extract relevant attributes and brands from the search query as well as from product titles. We then compare the performance of these three methods relative to each other.\\

\noindent We assume that there exists a bipartite graph $G: q \mapsto S$ that maps a customer's search query $q$ to a set of clicked SKUs $S$. This graph may be further augmented by including SKUs that were added to cart or bought. Search queries and SKUs are represented by nodes in the graph and an edge between a query and a SKU indicates that a customer searched for the query and clicked on the corresponding SKUs. The weight of the edge indicates the strength of the relationship between the query and the SKU and is modeled using number of clicks between the query and the SKU aggregated over a certain length of time. There are no edges between queries or between SKUs. Very broad queries like ``\texttt{cheap}'' or ``\texttt{clothing}'' either do not contain any products or contain very generic product terms and add noise to the data. We use entropy of a query across different categories to determine if it is broad and remove it from the graph. We also remove queries that are just brands from the graph and query-SKU pairs that have edge weights less than some threshold $(T)$. Finally, we apply a stemmer to perform stemming for terms in the query. Let $G'$ denote this cleaned bipartite graph. 

\noindent The task can be formulated as follows: Given a cleaned bipartite click graph $G'$, compute a sorted list of \emph{Product} sub-classes that are \emph{atomic} and \emph{relevant} for that category. We present three methods to create the sorted list of \emph{Product} classes and compare them.

\subsection{Token Graph Method}
\label{sec:token-graph-method}
\begin{figure}[!htbp]
  \includegraphics[width=\linewidth]{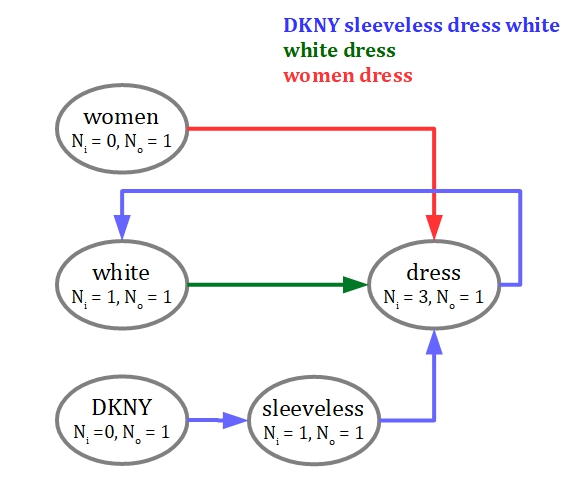}
  \caption{Token Graph Method.}
  \label{fig:graph_method_tokens}
\end{figure}

\noindent This method is a very simple unsupervised method for extracting relevant products from a customer's search query and can be applied to any category without any previous data. Let $C = \{q_0, q_1, \ldots, q_n, s_0, s_1, \ldots s_m\}$ be a connected component in the bipartite graph $G'$ mentioned previously. Let $Q = \{q_0, q_1, \ldots, q_n\}$ be a set of queries in this connected component and we can assume that all of them are related to each other because they share the same clicked SKUs. Let us assume that we can detect prepositions in the query and have removed them and all words after it from the query. Each token in the query is either a brand, product, attribute or other (part number, stopword, etc.) and we can create a new graph $G_{token}$ where each token is a node and there are edges between adjacent tokens. Figure~\ref{fig:graph_method_tokens} shows the token graph for the query set \emph{\{women dress, white dress, DKNY sleeveless dress white\}}. Most often, the product token is the last term in the query before any prepositions and thus it is the node that maximizes the ratio $\frac{N_i}{N_o + N_i}$, where $N_o$ is the number of outgoing edges and $N_i$ is the number of incoming edges for the node corresponding to the token. There are obvious exceptions to the rule, for example the search query: ``\texttt{DKNY sleeveless dress white}'' where the product \emph{dress} does not appear in the end of the query. However, we assume that such cases are rare and assume that aggregating this process over all related queries takes care of the occasional exception. We can generate a potential product from each connected component and aggregating over all connected components gives us a potential list of products.

\subsection{Augmented Graph Method}
\noindent The graph method in the previous section works pretty well but makes a very strong assumption that the product always appears towards the end of the search query. It is also very aggressive in removing the preposition and all tokens after it. For example, it will convert the query `'\texttt{seven for all mankind skinny jeans}'' to ``\texttt{seven}'', which is obviously wrong. Finally, it is oblivious to the parts-of-speech of the terms.\\

\noindent Typically, product words are nouns (television, shirt, etc.) and we can take advantage of parts-of-speech tags to improve the accuracy of the system. One option is to use global parts-of-speech tags from wordnet~\cite{miller1995wordnet} or some other similar repository. However, a word like \texttt{pack} may be used as a noun \texttt{(battery pack)} or a verb \texttt{(pack your stuff)} and the local context is lost if we use global parts-of-speech tags. Another problem with using a service like wordnet is that it may not contain some brand words like \texttt{Samsung}. A better approach is to use a service like Google's SyntaxNet~\cite{syntaxnet2016worlds} to generate parts-of-speech tags on the fly and this helps us retain local information as well as get parts-of-speech tags for brands like \texttt{Samsung}. We realized that most queries are not grammatically correct and so the generated parts-of-speech tags may not be very accurate. Instead, we ran SyntaxNet on the descriptions of all SKUs in $G'$ to generate a mapping between terms and their parts-of-speech tags. Let $\mathit{v^P_i = [NOUN,\,VERB,\,ADVERB,\,ADJ,\,PREP,\,NUM,\,\ldots]}$\\ denote a vector that represents the parts-of-speech for some term $t_i$. Here, $NOUN$ indicates the fraction of the time the part of speech tag for that term was a noun, $VERB$ indicates the fraction of the time the part of speech tag for that term was a verb and so on. We can use this map to generate parts-of-speech vectors for each term in the search query. \\

\noindent We also want to capture the local graph information discussed in the previous section. This can be done by creating the local graph and computing the number of incoming and outgoing edges for each term in the query. Let $v^G_i = [n_i,\,n_o,\,\frac{n_i}{n_i + n_o}]$ denote a vector that captures local graph information for the $i^{th}$ term. Here, $n_i$ indicates the number of incoming edges for the node denoting the term in the local graph and $n_o$ indicates the number of outgoing edges for the same node. If the search query contains just a single token, we set $n_i\,=n_o\,=1$ \\

\noindent Let $v^N_i\,=\,N\,-i$ denote a one-dimensional vector describing the position of the $i^{th}$ term in the search query, where $N$ is the number of terms in that query. This vector helps the model prefer later words in the query as products.\\

\noindent Finally, let $v_i=(v_i^P, v_i^G, v_i^N)$ denote a concatenated vector that captures all relevant information for the $i^{th}$ term in the query and let $V=(v_0,v_1, \ldots, v_n)$ denote the vector for the entire search term. We will use this vector as an input to the model to predict the product terms from the search query. We use a convolution neural network (CNN) that consists of three convolution layers with filter sizes of $n_1=7$ for the first layer, $n_2=5$ for the second layer and $n_3=3$ for the third layer. The number of filters are set to $256$ in each case. There is no max-pooling layer because we want to keep the filter information for each stride. The output of the last filter is then passed to fully-connected layers with a time-distributed-dense layer as the very last layer for making tag predictions. \\

\noindent The intuition behind this model is that the convolution layers are able to capture local information using the parts-of-speech tags of surrounding terms and the number of incoming and outgoing edges for the terms in the vicinity. It is then able to make a decision by combining all three vectors to predict if a term in the query is a product or not. The model is trained using queries across six categories (Electronics, Women's clothing, Men's clothing, Kid's clothing, Furniture, and Home) and the tested using queries from the Baby category. Each query can give zero or more product candidates and we aggregate candidates from all queries to come up with a list of potential products.

\subsection{NER Model using Bidirectional LSTM-CRF}
This model is very different from the two described earlier. It does not look at the local term graph but makes a decision using a word2vec~\cite{mikolov2013distributed} vector for each term in the query. The word2vec vectors are of dimension $D=300$ and are generated using data from Wikipedia and from SKU titles from the Jet.com catalog. The training data consists of queries where each term has been tagged in IOB format with either a O (other), B-PRODUCT (beginning of product) or I-PRODUCT (intermediate of product). For example, the query phrase \texttt{metal bar stool for kitchen} would be tagged as: \emph{metal O bar B-PRODUCT stool I-PPRODUCT for O kitchen O}. We use bi-directional LSTM-CRF model described in~\cite{glample} to train the NER model. The training data was tagged automatically using existing the existing query and SKU understanding service along with user engagement data to filter out potentially bad results.

\begin{equation}
\label{eq:lstm-crf}
\begin{split}
&h_t = o_t \odot \tanh(c_t)\\
&o_t = \sigma(W_{xo} x_t + W_{ho} h_{t-1} + W_{co} c_t + b_o)\\
&c_t = (1-i_t)\odot c_{t-1} + i_t\odot \tanh(W_{xc} x_t + W_{hc} h_{t-1} + b_c)\\
&i_t = \sigma(W_{xi} x_t + W_{hi} h_{t-1} + W_{ci} c_{t-1} + b_i)
\end{split}
\end{equation}\\

\noindent Let $S = (x_1,x_2,...,x_n)$ represent a sentence containing $n$ words where $x_t$ represents the word at position $t$ and each word is represented by a $d$-dimensional vector. We compute the left-context $\overrightarrow{h_t}$ using a forward LSTM and also a right-context $\overleftarrow{h_t}$ using a backward LSTM, which reads the same sequence in reverse order. The contexts $\overrightarrow{h_t}$ and $\overleftarrow{h_t}$ are computed as shown in equation~\ref{eq:lstm-crf}, where $\sigma$ is the element-wise sigmoid function, $\odot$ is the element-wise product, $W$ is the weight matrix and $b$ is the bias. The left and right contexts are then concatenated to represent a word representation $h_t=[\overrightarrow{h_t},\overleftarrow{h_t}]$, which is used by the conditional random field (CRF) for NER tagging.\\

\noindent Lexical features of queries can be quite different across categories. So for this method to generalize well, it was important to select the training dataset such that the labeled queries belonged to different categories. We chose queries from six categories (Electronics, Women's clothing, Men's clothing, Kid's clothing, Furniture, and Home) for training data and extracted candidate products using queries from the Baby category. 

\subsection{Model comparison}
The token graph method described in section~\ref{sec:token-graph-method} is an unsupervised model and so does not require any training data. The other two models are trained using labeled queries from six categories and all three models are tested using the same test set, which are queries from the Baby category. We exclude all broad queries and all queries that are just brands to keep it consistent with the training data. We believe that this is a fair test as it allows us evaluate the model's performance on a previously unseen category - a task that is essential for automatically creating ontologies.\\

\noindent Each model produces potential product candidates from queries and these candidates are sorted in decreasing order of frequency. We evaluate the top $500$ candidates from each model and manually verify if each potential product was actually a product or not. We consider a term to be a product only if it is \emph{atomic} and sellable on the site. For example, \emph{diaper} is a product but \emph{baby} (we don't sell babies) and \emph{diaper cover} (not atomic) are not. Table~\ref{tab:potential-products} shows the top ten candidates (from the top $500$ candidates) from each model along with our manually annotated results denoting if the given entry is a product $(P)$ or not $(N)$.\\

\begin{figure}[!htbp]
  \includegraphics[width=\linewidth]{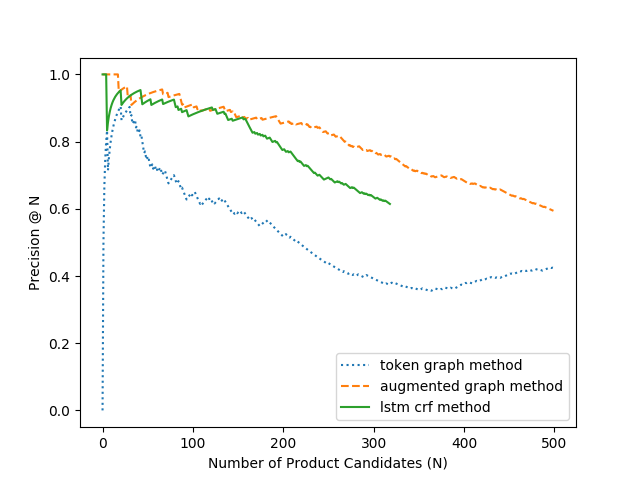}
  \caption{Precision @ N graphs for the three models (top 500 candidates).}
  \label{fig:comparison-500}
\end{figure}

\begin{figure}[!htbp]
  \includegraphics[width=\linewidth]{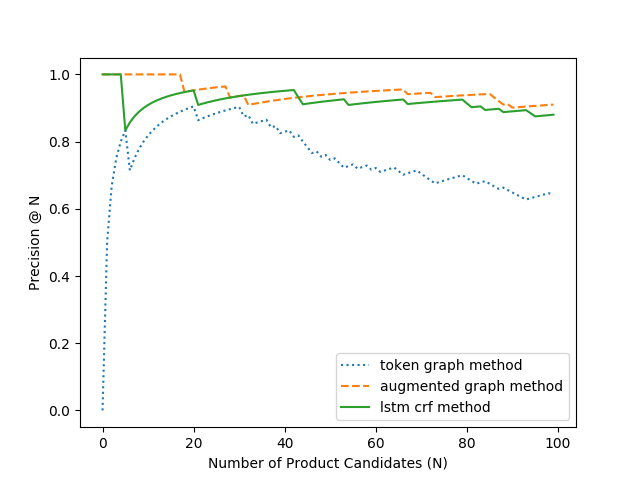}
  \caption{Precision @ N graphs for the three models (top 100 candidates).}
  \label{fig:comparison-100}
\end{figure}

\noindent Figure~\ref{fig:comparison-500} shows a \emph{precision @ n} graph for all the three models over their top $500$ candidates. The LSTM-CRF model produced just over $300$ candidates and so its graph is truncated. Both the augmented graph method and the LSTM-CRF method have a higher precision initially and are able to correctly identify products from the query logs. Figure~\ref{fig:comparison-100} shows a zoomed in view of the first $100$ candidates and it can be observed that the augmented graph model is able to predict products more accurately than the LSTM-CRF method. As expected the naive graph method performs the worst in terms of accuracy but has a better recall than the LSTM-CRF method.\\

\noindent The naive graph method may seem like the worst method but it has one very significant advantage over the other two methods - it is completely unsupervised. This allows it to be used when there is no training data from other categories. We recommend that this method should be used initially and it can pave the way for the other two supervised methods for other categories.

\begin{table}[!htbp]
\centering
\small
\caption{Top 10 Potential Products}
\label{tab:potential-products}
\begin{tabular}{|l|l|l|l|}
\hline
Num & Graph Method & Augmented Graph Model & NER Model \\ \hline
1   & all (N)        & diaper (P)             & diaper (P)        \\ \hline
2   & sippycup (P)    & wipe (P)                & wipe (P)          \\ \hline
3   & cup (P)         & formula (P)              & bottle (P)        \\ \hline
4   & bib (P)         & carseat (P)              & bag (P)           \\ \hline
5   & playard (P)    & bottle (P)               & cover (P)         \\ \hline
6   & insert (P)      & stroller (P)             & ups (N)           \\ \hline
7   & ct (N)          & bag (P)                  & pants (P)        \\ \hline
8   & highchair (P)  & gate (P)                 & seat (P)          \\ \hline
9   & case (P)       & cereal (P)               & pad (P)          \\ \hline
10  & stroller (P)   & highchair (P)            & bib (P)          \\ \hline
\end{tabular}
\end{table}

\section{Conclusion}
\label{sec:conclusion}
In this work we proposed a search-side ontology that can be used for Named Entity Recognition and Classification of Queries. We show that this ontology is better suited for search as compared to supply-side or catalog-side ontologies. We propose three methods to generate \emph{Product} classes for this ontology. We also compare the three methods and show that the Augmented Graph Method which uses local token information along with parts-of-speech tags performs better than the naive Graph Method and the bidirectional LSTM-CRF method in generating \emph{Product} classes.

\begin{acks}
The authors would like to thank Brent Scardapane and Dennis Jordan for their help with the ontology creation process.
\end{acks}

\bibliographystyle{ACM-Reference-Format}
\bibliography{sample-bibliography}


\begin{thebibliography}{28}


\ifx \showCODEN    \undefined \def \showCODEN     #1{\unskip}     \fi
\ifx \showDOI      \undefined \def \showDOI       #1{#1}\fi
\ifx \showISBNx    \undefined \def \showISBNx     #1{\unskip}     \fi
\ifx \showISBNxiii \undefined \def \showISBNxiii  #1{\unskip}     \fi
\ifx \showISSN     \undefined \def \showISSN      #1{\unskip}     \fi
\ifx \showLCCN     \undefined \def \showLCCN      #1{\unskip}     \fi
\ifx \shownote     \undefined \def \shownote      #1{#1}          \fi
\ifx \showarticletitle \undefined \def \showarticletitle #1{#1}   \fi
\ifx \showURL      \undefined \def \showURL       {\relax}        \fi
\providecommand\bibfield[2]{#2}
\providecommand\bibinfo[2]{#2}
\providecommand\natexlab[1]{#1}
\providecommand\showeprint[2][]{arXiv:#2}

\bibitem[\protect\citeauthoryear{Buitelaar and Sacaleanu}{Buitelaar and
  Sacaleanu}{2001}]%
        {buitelaar2001ranking}
\bibfield{author}{\bibinfo{person}{Paul Buitelaar} {and}
  \bibinfo{person}{Bogdan Sacaleanu}.} \bibinfo{year}{2001}\natexlab{}.
\newblock \showarticletitle{Ranking and selecting synsets by domain relevance}.
  In \bibinfo{booktitle}{\emph{Proceedings of WordNet and Other Lexical
  Resources: Applications, Extensions and Customizations, NAACL 2001
  Workshop}}. Citeseer, \bibinfo{pages}{119--124}.
\newblock


\bibitem[\protect\citeauthoryear{Charron, Hirate, Purcell, and Rezk}{Charron
  et~al\mbox{.}}{2016}]%
        {charron2016extracting}
\bibfield{author}{\bibinfo{person}{Bruno Charron}, \bibinfo{person}{Yu Hirate},
  \bibinfo{person}{David Purcell}, {and} \bibinfo{person}{Martin Rezk}.}
  \bibinfo{year}{2016}\natexlab{}.
\newblock \showarticletitle{Extracting semantic information for e-commerce}. In
  \bibinfo{booktitle}{\emph{International Semantic Web Conference}}. Springer,
  \bibinfo{pages}{273--290}.
\newblock


\bibitem[\protect\citeauthoryear{Cimiano}{Cimiano}{2006}]%
        {Cimiano:2006:OLP:1177318}
\bibfield{author}{\bibinfo{person}{Philipp Cimiano}.}
  \bibinfo{year}{2006}\natexlab{}.
\newblock \bibinfo{booktitle}{\emph{Ontology Learning and Population from Text:
  Algorithms, Evaluation and Applications}}.
\newblock \bibinfo{publisher}{Springer-Verlag}, \bibinfo{address}{Berlin,
  Heidelberg}.
\newblock
\showISBNx{0387306323}


\bibitem[\protect\citeauthoryear{Cunningham, Maynard, Bontcheva, and
  Tablan}{Cunningham et~al\mbox{.}}{2002}]%
        {Cunningham2002}
\bibfield{author}{\bibinfo{person}{Hamish Cunningham}, \bibinfo{person}{Diana
  Maynard}, \bibinfo{person}{Kalina Bontcheva}, {and} \bibinfo{person}{Valentin
  Tablan}.} \bibinfo{year}{2002}\natexlab{}.
\newblock \showarticletitle{{GATE: A Framework and Graphical Development
  Environment for Robust NLP Tools and Applications}}. In
  \bibinfo{booktitle}{\emph{Proceedings of the 40th Anniversary Meeting of the
  Association for Computational Linguistics (ACL'02)}}.
\newblock


\bibitem[\protect\citeauthoryear{Cunningham, Tablan, Roberts, and
  Bontcheva}{Cunningham et~al\mbox{.}}{2013}]%
        {cunningham2013getting}
\bibfield{author}{\bibinfo{person}{Hamish Cunningham},
  \bibinfo{person}{Valentin Tablan}, \bibinfo{person}{Angus Roberts}, {and}
  \bibinfo{person}{Kalina Bontcheva}.} \bibinfo{year}{2013}\natexlab{}.
\newblock \showarticletitle{Getting more out of biomedical documents with
  GATE's full lifecycle open source text analytics}.
\newblock \bibinfo{journal}{\emph{PLoS computational biology}}
  \bibinfo{volume}{9}, \bibinfo{number}{2} (\bibinfo{year}{2013}),
  \bibinfo{pages}{e1002854}.
\newblock


\bibitem[\protect\citeauthoryear{Ding, Fensel, Klein, Omelayenko, and
  Schulten}{Ding et~al\mbox{.}}{2004}]%
        {ding2004role}
\bibfield{author}{\bibinfo{person}{Ying Ding}, \bibinfo{person}{Dieter Fensel},
  \bibinfo{person}{Michel Klein}, \bibinfo{person}{Borys Omelayenko}, {and}
  \bibinfo{person}{Ellen Schulten}.} \bibinfo{year}{2004}\natexlab{}.
\newblock \showarticletitle{The role of ontologies in ecommerce}.
\newblock In \bibinfo{booktitle}{\emph{Handbook on ontologies}}.
  \bibinfo{publisher}{Springer}, \bibinfo{pages}{593--615}.
\newblock


\bibitem[\protect\citeauthoryear{Frantzi and Ananiadou}{Frantzi and
  Ananiadou}{1999}]%
        {frantzi1999c}
\bibfield{author}{\bibinfo{person}{Katerina~T Frantzi} {and}
  \bibinfo{person}{Sophia Ananiadou}.} \bibinfo{year}{1999}\natexlab{}.
\newblock \showarticletitle{The C-value/NC-value domain-independent method for
  multi-word term extraction}.
\newblock \bibinfo{journal}{\emph{Journal of Natural Language Processing}}
  \bibinfo{volume}{6}, \bibinfo{number}{3} (\bibinfo{year}{1999}),
  \bibinfo{pages}{145--179}.
\newblock


\bibitem[\protect\citeauthoryear{Gabrilovich and Markovitch}{Gabrilovich and
  Markovitch}{2007}]%
        {gabrilovich2007computing}
\bibfield{author}{\bibinfo{person}{Evgeniy Gabrilovich} {and}
  \bibinfo{person}{Shaul Markovitch}.} \bibinfo{year}{2007}\natexlab{}.
\newblock \showarticletitle{Computing semantic relatedness using
  wikipedia-based explicit semantic analysis.}. In
  \bibinfo{booktitle}{\emph{IJcAI}}, Vol.~\bibinfo{volume}{7}.
  \bibinfo{pages}{1606--1611}.
\newblock


\bibitem[\protect\citeauthoryear{Glater, Santos, and Ziviani}{Glater
  et~al\mbox{.}}{2017}]%
        {glater2017intent}
\bibfield{author}{\bibinfo{person}{Rafael Glater}, \bibinfo{person}{Rodrygo~LT
  Santos}, {and} \bibinfo{person}{Nivio Ziviani}.}
  \bibinfo{year}{2017}\natexlab{}.
\newblock \showarticletitle{Intent-Aware Semantic Query Annotation}. In
  \bibinfo{booktitle}{\emph{Proceedings of the 40th International ACM SIGIR
  Conference on Research and Development in Information Retrieval}}. ACM,
  \bibinfo{pages}{485--494}.
\newblock


\bibitem[\protect\citeauthoryear{Gormley and Tong}{Gormley and Tong}{2015}]%
        {gormley2015elasticsearch}
\bibfield{author}{\bibinfo{person}{Clinton Gormley} {and}
  \bibinfo{person}{Zachary Tong}.} \bibinfo{year}{2015}\natexlab{}.
\newblock \bibinfo{booktitle}{\emph{Elasticsearch: The Definitive Guide: A
  Distributed Real-Time Search and Analytics Engine}}.
\newblock \bibinfo{publisher}{" O'Reilly Media, Inc."}.
\newblock


\bibitem[\protect\citeauthoryear{Grainger, Potter, and Seeley}{Grainger
  et~al\mbox{.}}{2014}]%
        {grainger2014solr}
\bibfield{author}{\bibinfo{person}{Trey Grainger}, \bibinfo{person}{Timothy
  Potter}, {and} \bibinfo{person}{Yonik Seeley}.}
  \bibinfo{year}{2014}\natexlab{}.
\newblock \bibinfo{booktitle}{\emph{Solr in action}}.
\newblock \bibinfo{publisher}{Manning Cherry Hill}.
\newblock


\bibitem[\protect\citeauthoryear{Hearst}{Hearst}{1992}]%
        {hearst1992automatic}
\bibfield{author}{\bibinfo{person}{Marti~A Hearst}.}
  \bibinfo{year}{1992}\natexlab{}.
\newblock \showarticletitle{Automatic acquisition of hyponyms from large text
  corpora}. In \bibinfo{booktitle}{\emph{Proceedings of the 14th conference on
  Computational linguistics-Volume 2}}. Association for Computational
  Linguistics, \bibinfo{pages}{539--545}.
\newblock


\bibitem[\protect\citeauthoryear{Hu, Wang, Lochovsky, Sun, and Chen}{Hu
  et~al\mbox{.}}{2009}]%
        {hu2009understanding}
\bibfield{author}{\bibinfo{person}{Jian Hu}, \bibinfo{person}{Gang Wang},
  \bibinfo{person}{Fred Lochovsky}, \bibinfo{person}{Jian-tao Sun}, {and}
  \bibinfo{person}{Zheng Chen}.} \bibinfo{year}{2009}\natexlab{}.
\newblock \showarticletitle{Understanding user's query intent with wikipedia}.
  In \bibinfo{booktitle}{\emph{Proceedings of the 18th international conference
  on World wide web}}. ACM, \bibinfo{pages}{471--480}.
\newblock


\bibitem[\protect\citeauthoryear{Lample, Ballesteros, Subramanian, Kawakami,
  and Dyer}{Lample et~al\mbox{.}}{2016}]%
        {glample}
\bibfield{author}{\bibinfo{person}{Guillaume Lample}, \bibinfo{person}{Miguel
  Ballesteros}, \bibinfo{person}{Sandeep Subramanian}, \bibinfo{person}{Kazuya
  Kawakami}, {and} \bibinfo{person}{Chris Dyer}.}
  \bibinfo{year}{2016}\natexlab{}.
\newblock \showarticletitle{Neural Architectures for Named Entity Recognition}.
\newblock \bibinfo{journal}{\emph{CoRR}}  \bibinfo{volume}{abs/1603.01360}
  (\bibinfo{year}{2016}).
\newblock
\showeprint[arxiv]{1603.01360}
\urldef\tempurl%
\url{http://arxiv.org/abs/1603.01360}
\showURL{%
\tempurl}


\bibitem[\protect\citeauthoryear{Lee, Lee, Lee, Lee, Kim, Chun, Lee, and
  Shim}{Lee et~al\mbox{.}}{2006}]%
        {lee2006building}
\bibfield{author}{\bibinfo{person}{Taehee Lee}, \bibinfo{person}{Ig-hoon Lee},
  \bibinfo{person}{Suekyung Lee}, \bibinfo{person}{Sang-goo Lee},
  \bibinfo{person}{Dongkyu Kim}, \bibinfo{person}{Jonghoon Chun},
  \bibinfo{person}{Hyunja Lee}, {and} \bibinfo{person}{Junho Shim}.}
  \bibinfo{year}{2006}\natexlab{}.
\newblock \showarticletitle{Building an operational product ontology system}.
\newblock \bibinfo{journal}{\emph{Electronic Commerce Research and
  Applications}} \bibinfo{volume}{5}, \bibinfo{number}{1}
  (\bibinfo{year}{2006}), \bibinfo{pages}{16--28}.
\newblock


\bibitem[\protect\citeauthoryear{Maedche and Staab}{Maedche and Staab}{2000}]%
        {maedche2000discovering}
\bibfield{author}{\bibinfo{person}{Alexander Maedche} {and}
  \bibinfo{person}{Steffen Staab}.} \bibinfo{year}{2000}\natexlab{}.
\newblock \showarticletitle{Discovering conceptual relations from text}. In
  \bibinfo{booktitle}{\emph{Ecai}}, Vol.~\bibinfo{volume}{321}.
  \bibinfo{pages}{27}.
\newblock


\bibitem[\protect\citeauthoryear{Maedche and Staab}{Maedche and Staab}{2004}]%
        {maedche2004ontology}
\bibfield{author}{\bibinfo{person}{Alexander Maedche} {and}
  \bibinfo{person}{Steffen Staab}.} \bibinfo{year}{2004}\natexlab{}.
\newblock \showarticletitle{Ontology learning}.
\newblock In \bibinfo{booktitle}{\emph{Handbook on ontologies}}.
  \bibinfo{publisher}{Springer}, \bibinfo{pages}{173--190}.
\newblock


\bibitem[\protect\citeauthoryear{Mikolov, Sutskever, Chen, Corrado, and
  Dean}{Mikolov et~al\mbox{.}}{2013}]%
        {mikolov2013distributed}
\bibfield{author}{\bibinfo{person}{Tomas Mikolov}, \bibinfo{person}{Ilya
  Sutskever}, \bibinfo{person}{Kai Chen}, \bibinfo{person}{Greg~S Corrado},
  {and} \bibinfo{person}{Jeff Dean}.} \bibinfo{year}{2013}\natexlab{}.
\newblock \showarticletitle{Distributed representations of words and phrases
  and their compositionality}. In \bibinfo{booktitle}{\emph{Advances in neural
  information processing systems}}. \bibinfo{pages}{3111--3119}.
\newblock


\bibitem[\protect\citeauthoryear{Miller}{Miller}{1995}]%
        {miller1995wordnet}
\bibfield{author}{\bibinfo{person}{George~A Miller}.}
  \bibinfo{year}{1995}\natexlab{}.
\newblock \showarticletitle{WordNet: a lexical database for English}.
\newblock \bibinfo{journal}{\emph{Commun. ACM}} \bibinfo{volume}{38},
  \bibinfo{number}{11} (\bibinfo{year}{1995}), \bibinfo{pages}{39--41}.
\newblock


\bibitem[\protect\citeauthoryear{Nadeau and Sekine}{Nadeau and Sekine}{2007}]%
        {nadeau2007survey}
\bibfield{author}{\bibinfo{person}{David Nadeau} {and} \bibinfo{person}{Satoshi
  Sekine}.} \bibinfo{year}{2007}\natexlab{}.
\newblock \showarticletitle{A survey of named entity recognition and
  classification}.
\newblock \bibinfo{journal}{\emph{Lingvisticae Investigationes}}
  \bibinfo{volume}{30}, \bibinfo{number}{1} (\bibinfo{year}{2007}),
  \bibinfo{pages}{3--26}.
\newblock


\bibitem[\protect\citeauthoryear{Noy, McGuinness, et~al\mbox{.}}{Noy
  et~al\mbox{.}}{2001}]%
        {noy2001ontology}
\bibfield{author}{\bibinfo{person}{Natalya~F Noy}, \bibinfo{person}{Deborah~L
  McGuinness}, {et~al\mbox{.}}} \bibinfo{year}{2001}\natexlab{}.
\newblock \bibinfo{title}{Ontology development 101: A guide to creating your
  first ontology}.
\newblock
\newblock


\bibitem[\protect\citeauthoryear{Petrucci, Ghidini, and Rospocher}{Petrucci
  et~al\mbox{.}}{2016}]%
        {petrucci2016ontology}
\bibfield{author}{\bibinfo{person}{Giulio Petrucci}, \bibinfo{person}{Chiara
  Ghidini}, {and} \bibinfo{person}{Marco Rospocher}.}
  \bibinfo{year}{2016}\natexlab{}.
\newblock \showarticletitle{Ontology learning in the deep}. In
  \bibinfo{booktitle}{\emph{European Knowledge Acquisition Workshop}}.
  Springer, \bibinfo{pages}{480--495}.
\newblock


\bibitem[\protect\citeauthoryear{Popov, Kiryakov, Ognyanoff, Manov, and
  Kirilov}{Popov et~al\mbox{.}}{2004}]%
        {popov2004kim}
\bibfield{author}{\bibinfo{person}{Borislav Popov}, \bibinfo{person}{Atanas
  Kiryakov}, \bibinfo{person}{Damyan Ognyanoff}, \bibinfo{person}{Dimitar
  Manov}, {and} \bibinfo{person}{Angel Kirilov}.}
  \bibinfo{year}{2004}\natexlab{}.
\newblock \showarticletitle{KIM--a semantic platform for information extraction
  and retrieval}.
\newblock \bibinfo{journal}{\emph{Natural language engineering}}
  \bibinfo{volume}{10}, \bibinfo{number}{3-4} (\bibinfo{year}{2004}),
  \bibinfo{pages}{375--392}.
\newblock


\bibitem[\protect\citeauthoryear{Rose and Levinson}{Rose and Levinson}{2004}]%
        {rose2004understanding}
\bibfield{author}{\bibinfo{person}{Daniel~E Rose} {and} \bibinfo{person}{Danny
  Levinson}.} \bibinfo{year}{2004}\natexlab{}.
\newblock \showarticletitle{Understanding user goals in web search}. In
  \bibinfo{booktitle}{\emph{Proceedings of the 13th international conference on
  World Wide Web}}. ACM, \bibinfo{pages}{13--19}.
\newblock


\bibitem[\protect\citeauthoryear{S{\'a}nchez and Moreno}{S{\'a}nchez and
  Moreno}{2005}]%
        {sanchez2005web}
\bibfield{author}{\bibinfo{person}{David S{\'a}nchez} {and}
  \bibinfo{person}{Antonio Moreno}.} \bibinfo{year}{2005}\natexlab{}.
\newblock \showarticletitle{Web-scale taxonomy learning}. In
  \bibinfo{booktitle}{\emph{Proceedings of Workshop on Extending and Learning
  Lexical Ontologies using Machine Learning (ICML 2005)}}.
  \bibinfo{pages}{53--60}.
\newblock


\bibitem[\protect\citeauthoryear{SyntaxNet}{SyntaxNet}{2016}]%
        {syntaxnet2016worlds}
\bibfield{author}{\bibinfo{person}{Announcing SyntaxNet}.}
  \bibinfo{year}{2016}\natexlab{}.
\newblock \bibinfo{title}{The Worlds Most Accurate Parser Goes Open Source}.
\newblock
\newblock


\bibitem[\protect\citeauthoryear{Wohlgenannt and Minic}{Wohlgenannt and
  Minic}{2016}]%
        {wohlgenannt2016using}
\bibfield{author}{\bibinfo{person}{Gerhard Wohlgenannt} {and}
  \bibinfo{person}{Filip Minic}.} \bibinfo{year}{2016}\natexlab{}.
\newblock \showarticletitle{Using word2vec to Build a Simple Ontology Learning
  System.}. In \bibinfo{booktitle}{\emph{International Semantic Web Conference
  (Posters \& Demos)}}.
\newblock


\bibitem[\protect\citeauthoryear{Zhang, Iria, Brewster, and Ciravegna}{Zhang
  et~al\mbox{.}}{2008}]%
        {zhang2008comparative}
\bibfield{author}{\bibinfo{person}{Ziqi Zhang}, \bibinfo{person}{Jos{\'e}
  Iria}, \bibinfo{person}{Christopher Brewster}, {and} \bibinfo{person}{Fabio
  Ciravegna}.} \bibinfo{year}{2008}\natexlab{}.
\newblock \showarticletitle{A comparative evaluation of term recognition
  algorithms}.
\newblock  (\bibinfo{year}{2008}).
\newblock


\end{thebibliography}

\end{document}